# Band gap studies of nanocomposites of ZnO/SnO$_2$ with different molar ratios

*J.V.S.S.D. Perera*

Department of Physics, University of Peradeniya, Peradeniya, Sri Lanka

Abstract

Nano particles of pure ZnO, pure SnO$_2$, and nanocomposites of ZnO/SnO$_2$ were synthesized using microwave hydrothermal technique starting from aqueous solutions. Nanocomposites with different molar ratios of ZnO and SnO$_2$ were prepared. Structural, morphological and optical properties of these samples were investigated using XRD, SEM and UV-visible technique, respectively. The single phases of ZnO and SnO$_2$ could be fabricated according to XRD patterns. The relative intensities of different XRD peaks in different nanocomposite samples were different by indicating that there are some preferred orientations in different samples. In XRD patterns of nanocomposites, peaks of both pure ZnO and pure SnO$_2$ samples appear. Particle size was in the range of nano-range according to SEM micrographs. Optical bang gap was measured by UV-visible spectrometer. Optical band gaps of nano composites are less than those of pure ZnO and SnO$_2$ due to the increase of particle size.

## 1. Introduction:

ZnO is transparent to visible light owing to its wide band gap of 3.21 eV. ZnO are used in photovoltaic applications, gas sensors, antibacterial treatments, sunscreen lotions, photocatalysis and biological applications. Crystal structure of SnO$_2$ is tetragonal, and having band gap 3.78 eV. SnO$_2$ find potential applications in photocatalysis, gas sensors, solar cells and lithium ion batteries. Owing to low cost and easy preparation methods, metal oxide are utilized in many applications. Synthesis techniques of hierarchical SnO$_2$ nanostructures and their applications have been widely explained [1]. Synthesis and applications of nanowires, nanobelts and nanotubes of SnO$_2$ have been described [2]. Preparation of SnO$_2$ nanoparticles using biological based techniques have been presented [3]. SnO$_2$ nanoparticles have been synthesized by controlling particle size and distribution using Nb$_2$O$_5$ additive in a polymeric precursor method. According to these studies, Nb$_2$O$_5$ doped SnO$_2$ respond to gas faster than undoped SnO$_2$ nanoparticle [4]. Photocatalytic degradation of $SnO_2$ nanorods synthesized in polyethylene glycol medium measured in Rhodamine B has reached 100% in 45 minutes under irradiation of 300 W high pressure mercury lamp [5].

ZnO is employed in organic and hybrid solar cells. In these solar cells, ZnO and organic semiconductor serve as electron acceptor and electron donor materials, respectively [6]. Photocatalysis applications of ZnO nanowires have been presented [7]. Photovoltaic properties of DC sputtered ZnO has been studied [8]. Sputtered ZnO films can be employed as $CO_2$ gas sensors [9]. Photovoltaic properties of ZnO films can be improved by coating dye [10]. Optical and electrical properties of Cu doped ZnO nanocomposite films have been investigated [11, 12]. Photovoltaic properties of bilayer CuO/ZnO films have been studied [13]. Most of the oxide materials find potential applications in gas sensors. [14]. These kinds of oxide nanocomposites are used as photocatalysis. Photocatalytic properties of titania-silica mixed oxide mesoporous materials has been investigated by controlling pH value of acidic media [15]. Photocatalytic properties of spheres of titanium dioxide, zirconium dioxide, iron oxide, aluminum oxide, indium oxide, tin oxide and cerium oxide were studied [16]. Photocatalytic properties of zirconium oxide-zinc oxide nanoparticles fabricated using microwave irradiation have been investigated [17]. The highest photocatalytic efficiency obtained for this compound was 97%. Optoelectronic and photocatalytic properties of ZnO and $V_2O_5$ have been studied [18]. Photocatalytic properties of TiO2/Fe2O3 mixed oxides have been investigated [19].

## 2. Experimental:

Zinc Nitrate ($Zn(NO_3)_2.6H_2O$), Tin chloride ($SnCl_2.2H_2O$), Ammonium Hydroxide, were used without further purification. Methylene Blue ($C_{16}H_{18}N_3SCl.3H_2O$, Sigma Aldrich), Absolute ethanol (98%, Sigma Aldrich) was used as a solvent in the cleansing processes. Distilled deionized water was used to prepare all the required solutions.

0.1 mol $L^{-1}$ of $Zn(NO_3)_2.6H_2O$ solution was prepared, and the solution was stirred for 10 minutes to prepare ZnO nanoparticles. This solution was mixed with 5 ml of ammonia solution to prepare a precipitating agent. After stirring this solution for 20 minutes, it was irradiated in the microwave oven in 600 W for 10 minutes. After the precipitate came to the room temperature, powder was carefully separated and washed with deionized water and absolute ethanol. Thereafter, the powder was dried at 120 ºC for an hour. Then, ZnO nanoparticles were annealed at 800 ºC for 5 hours in air, and labeled as Z1. Similar approach was employed to fabricate $SnO_2$ nanoparticles by starting with a 0.1 mol L$^{-1}$ SnCl$_2$.2H$_2$O, and the sample was named as S2. Following the similar experimental method, nano composite samples were made with Zn:Sn molar ratios of 1:1, 1:2, 2:1, and the samples were labeled as ZS1, ZS2, ZS3, respectively.

Absorption properties and optical bandgap of powder samples were measured by means of a Shimadzu 1800 UV-Vis spectrophotometer. Structural properties of samples were investigated using a Rigaku Ultima IV X-Ray diffractometer with Cu-Kα (λ=1.5406 Å) radiation. Surface morphology of the samples were determined using a Zeiss EVO LS15 Scanning Electron Microscope. Photocatalytic degradation was investigated using a Vernier Go Direct SpectroVis Plus.

## 3. Results and discussion:

Figure 1 shows the XRD patterns of pure ZnO (black line), pure SnO$_2$ (red), ZnO:SnO$_2$ = 1:1 (blue line), ZnO:SnO$_2$ = 1:2 (green line), and ZnO:SnO$_2$ = 2:1 (purple line) nanocomposite samples. Narrow XRD peaks confirm the formation of nanoparticles. According to XRD patterns, only the phases of ZnO and SnO$_2$ were found in nanocomposite samples. Formation of any secondary phases due to the mixing of ZnO and SnO$_2$ was not found. The relative intensities of XRD peaks of nanocomposites can be used to find the ratio between ZnO and SnO$_2$. Some peaks of pure ZnO can not be found in XRD pattern of nanocomposite of ZnO:SnO$_2$ = 1:2 due to the less amount of ZnO in the composite.

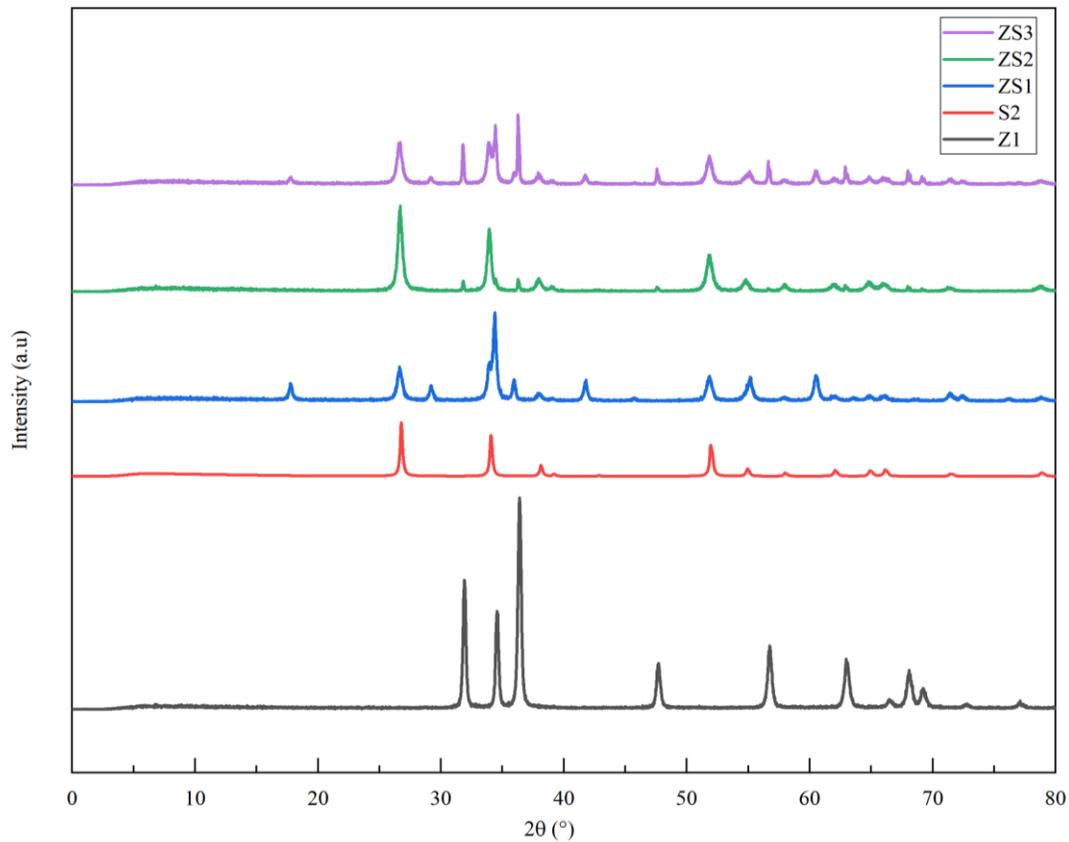

**Fig. 1** XRD patterns for samples of pure ZnO (black line), pure SnO$_2$ (red), ZnO:SnO$_2$ = 1:1 (blue line), ZnO:SnO$_2$ = 1:2 (green line), and ZnO:SnO$_2$ = 2:1 (purple line).

Figure 2 represents the SEM micrographs of pure ZnO and SnO$_2$ samples. This micrograph also verifies the formation of nanoparticles. The particle size of pure SnO$_2$ is smaller than the particle size of pure ZnO. The particles of nanocomposites are also in the nano-range.

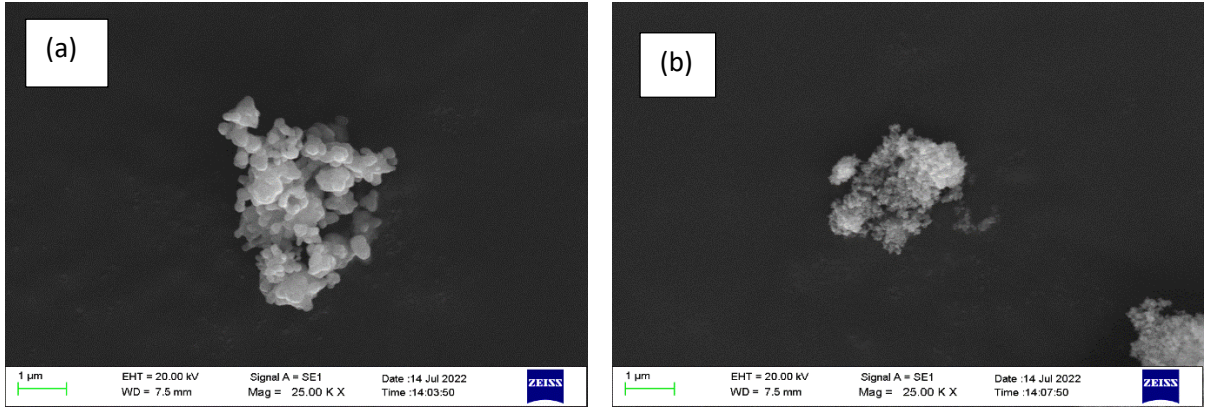

**Fig. 2** SEM images of the nano particle samples (a) pure ZnO (b) pure SnO$_2$ at a 25.00k magnification

In crystalline semiconductors, to relate the absorption coefficient to incident photon energy for direct transitions, the following equation is used.

$$\alpha h\nu = B(h\nu - E_g)^m \tag{1}$$

Where,

- $B$ - Constant related to effective mass of electrons and holes.
- $E_g$ - Optical band gap energy
- $h$ - incident photon energy
- $\alpha$ - absorption coefficient

Rewriting equation 1,

$$\alpha h\nu = B(hc)^{m-1} \lambda \left(\frac{1}{\lambda} - \frac{1}{\lambda_g}\right)^m \tag{2}$$

where,

- $\lambda_g$ - wavelength corresponding to the optical band gap
- $c$ - velocity of light

Using Beer Lambert's law;

$$A = B_1 \lambda \left(\frac{1}{\lambda} - \frac{1}{\lambda_g}\right)^m + B_2 \tag{3}$$

where,

$B_1 = B(hc)^{m-1} \times \frac{l}{2.303}$

$B_2$ - constant that take in to account the reflection.

$l$ - thickness of the film

Using equation 3, the optical band gap can be calculated by an absorbance spectrum fitting method without the film thickness. The value of $\lambda_g$ can be obtained by extrapolating the linear region of the $\left(\frac{A}{\lambda}\right)^{\frac{1}{m}}$ vs $\left(\frac{1}{\lambda}\right)$ curve at $\left(\frac{A}{\lambda}\right)^{\frac{1}{m}} = 0$.

By using the least squares technique, it has been observed that the best fitting occurs for m = 1/2. The $\lambda_g$ value obtained from the graph can be used to calculate the value of $E_g$.

$$E_g = \frac{hc}{\lambda} \qquad (4)$$

Figure 3 shows the UV-Visible absorption spectra of ZnO and SnO$_2$ nanoparticle samples. Figure 4 represents the absorption spectra of the ZnO:SnO$_2$ = 1:1 (black line), ZnO:SnO$_2$ = 1:2 (red line), and ZnO:SnO$_2$ = 2:1 (blue line) nanocomposites.

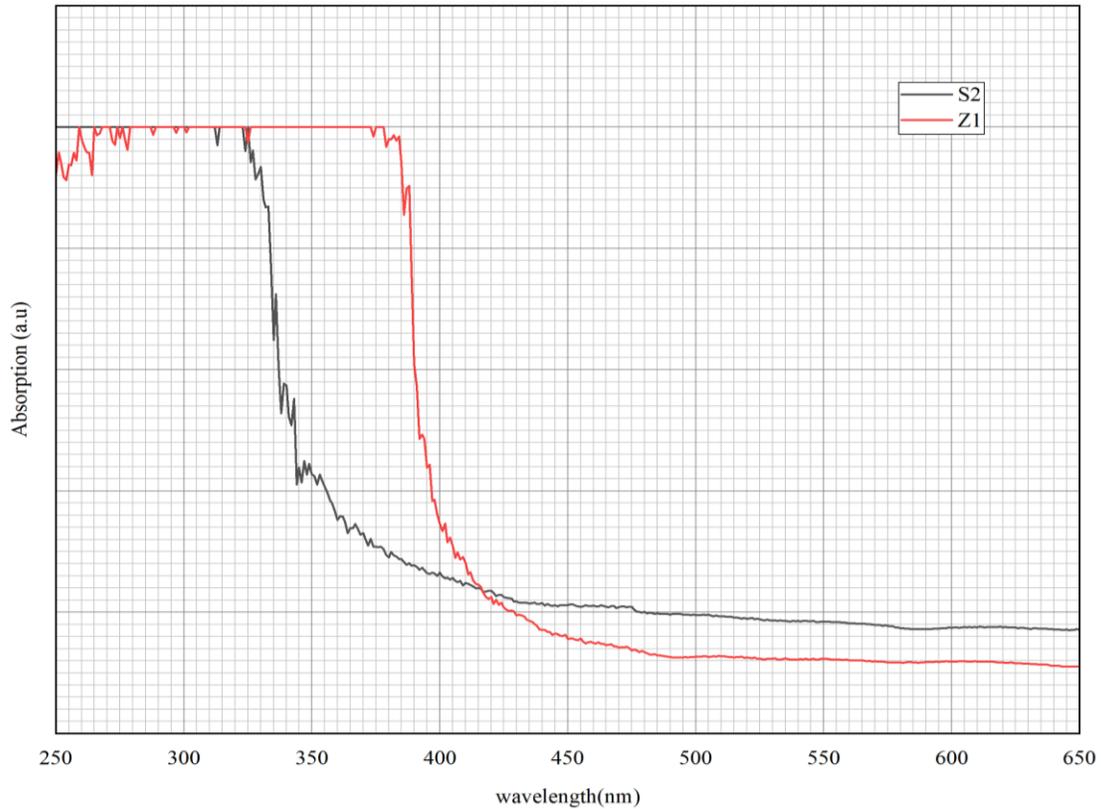

**Fig. 3** Absorption spectra of the pure ZnO and SnO$_2$ samples in UV-Vis analysis

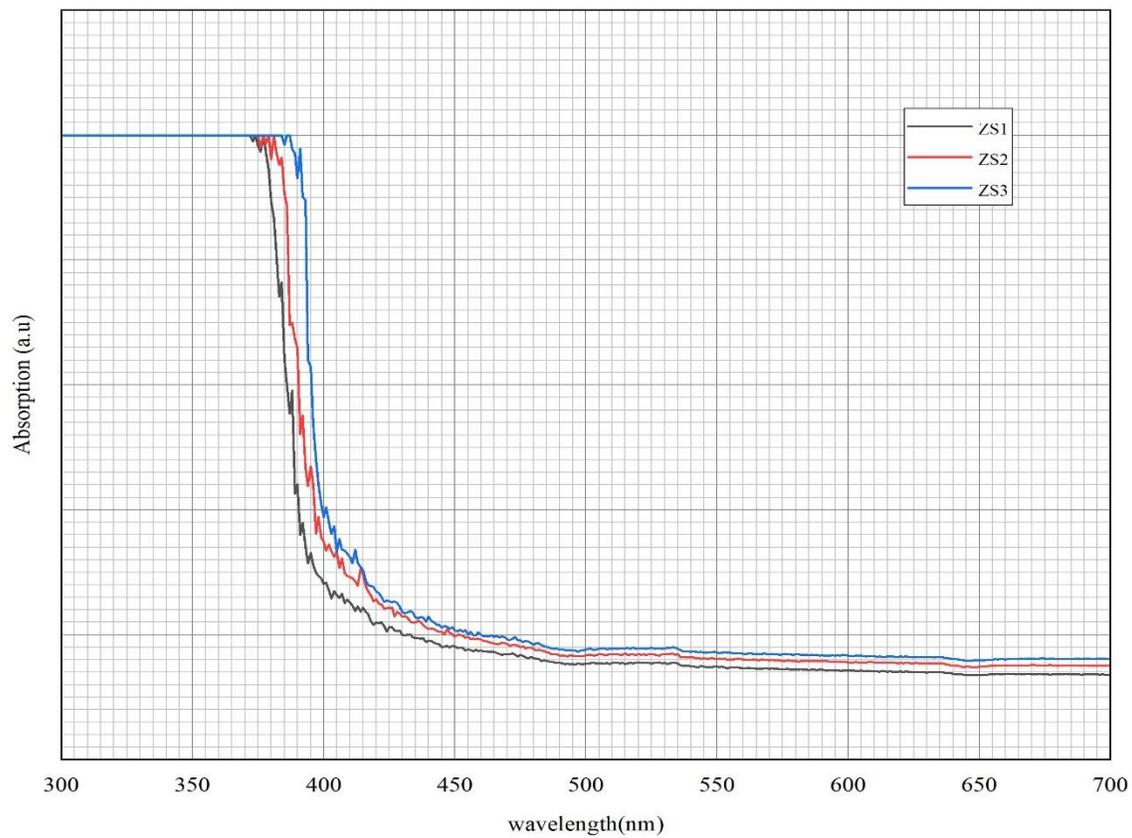

**Fig. 4** Absorption spectra of the ZnO:SnO$_2$ = 1:1 (black line), ZnO:SnO$_2$ = 1:2 (red line), and ZnO:SnO$_2$ = 2:1 (blue line).

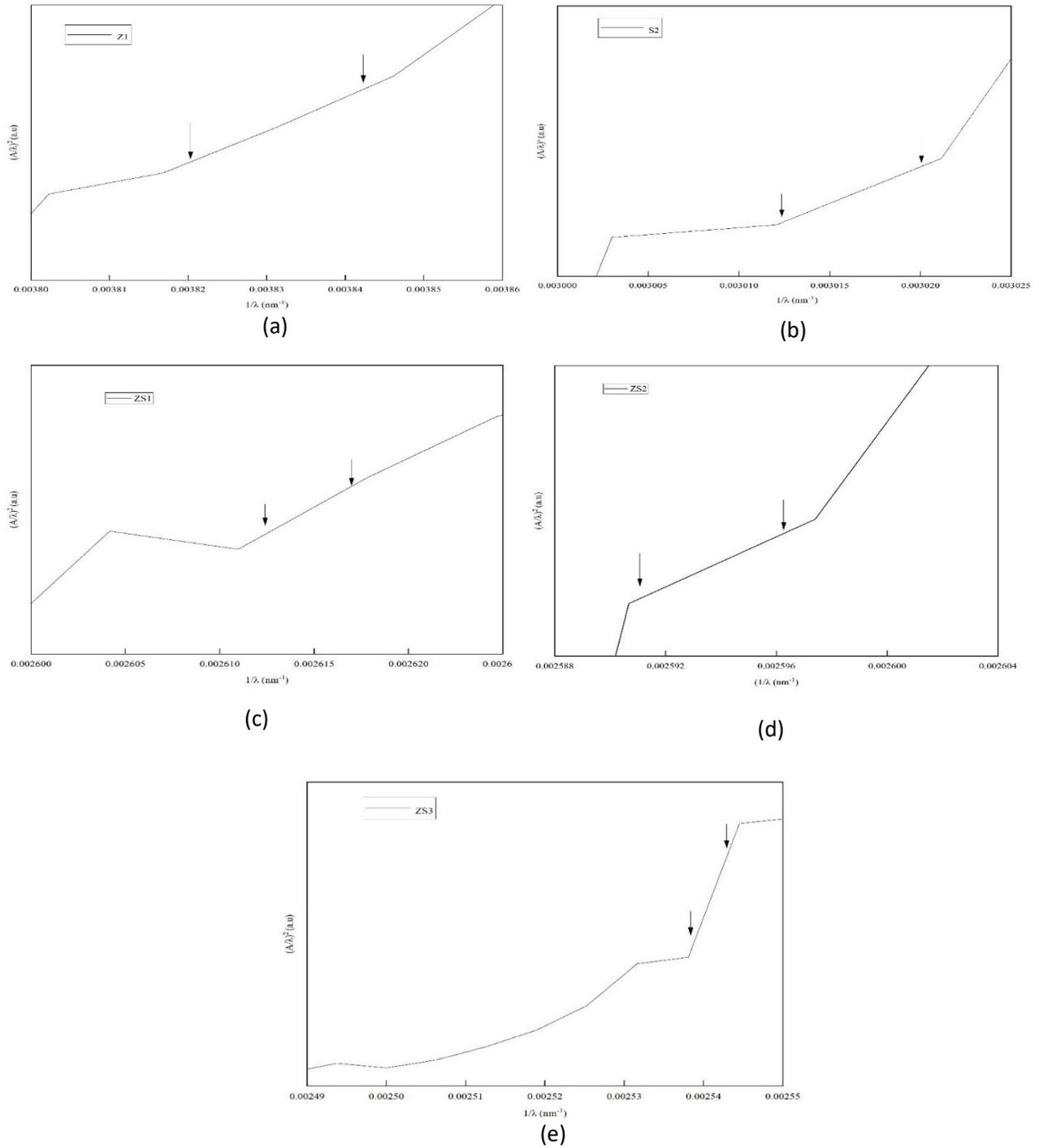

**Fig. 5** Graphs of $(A/\lambda)1/m$ vs. $(1/\lambda)$ for nanoparticle samples of (a) pure ZnO (b) pure SnO$_2$ (c) ZnO:SnO$_2$ = 1:1 (d) ZnO:SnO$_2$ = 1:2 and (e) ZnO:SnO$_2$ = 2:1

A linear behavior within certain region of all plots of $(A/\lambda)^{1/m}$ vs. $(1/\lambda)$ for different samples was observed, which corresponds to the region between the absorption edge and absorption peak of the absorbance spectra obtained from UV-Visible spectroscopy. These linear regions were extrapolated x-axis to calculate the respective bandgaps of prepared samples.

The calculated band gaps for each sample and the respective wavelengths absorbed by the band gaps are given in table 1.

**Table 1:** Calculated band gaps for each prepared nanoparticle

| Sample | $E_g$ (eV) | $\lambda$ (x$10^{-9}$ m) |
|---|---|---|
| Pure ZnO | 3.229 | 384 |
| Pure SnO$_2$ | 3.724 | 333 |
| ZnO:SnO$_2$ = 1:1 | 3.221 | 385 |
| ZnO:SnO$_2$ = 1:2 | 3.204 | 387 |
| ZnO:SnO$_2$ = 2:1 | 3.155 | 395 |

According to the obtained results, energy band gap of all the nanocomposite samples are less compared to band gap values of pure ZnO and SnO$_2$. The accumulation of particles in the mixing process is attributed to the larger particle size in nanocomposites. According to the Brus equation, the energy gap decreases as the particle size increases.

**4. Conclusion:**

According to SEM micrograph, the particle size of pure SnO$_2$ is smaller than the particle size of pure ZnO. Some XRD peaks of pure ZnO is not visible in XRD pattern of nanocomposite of ZnO:SnO$_2$ = 1:2 due to the less amount of ZnO in the composite. UV-Vis analysis resulted with 3.22 and 3.72 eV band gap energies for pure ZnO and SnO$_2$, respectively, and 3.22, 3.20 and 3.15 eV band gap energies for 1:1, 1:2 and 2:1 molar ratios of Zn:Sn nanocomposites, respectively. The optical bang reduces after mixing of pure ZnO and SnO$_2$. Below 450 nm wavelength, the absorption of pure ZnO is higher compared to pure SnO$_2$. Above 450 nm, the absorption of pure ZnO is less compared to pure SnO$_2$. Below wavelength of 350 nm, both samples have almost the same absorption. Absorption of ZnO:$SnO_2$ = 2:1 nanocomposite is higher compared to the other two nanocomposites. The sample with the higher amount of ZnO indicates the higher absorption. This can be attributed to the lower band gap of pure ZnO compared to pure $SnO_2$.

**References:**

1. H. Wang and A.L. Rogach. Hierarchical $SnO_2$ nanostructures: Recent advances in design, synthesis and applications. *Chemistry of Msterials* 2014; 26(1): 123-133.
2. J. Pan, H. Shen and S. Mathur. One dimensional SnO2 nanostructures: Synthesis and their applications. *Journal of Nanotechnology* 2012; 2012: Article ID 917320.
3. S Gorai. Bio-based synthesis and applications of $SnO_2$ nanoparticles- An overview. *Journal of Materials and Environmental Sciences* 2018; 9(10): 2894-2903.
4. E.R. Leite, I.T. Weber, E. Longo, J.A. Varela. A new method to control particle size and particle size distribution of $SnO_2$ nanoparticles for gas sensor applications. *Advanced Materials* 2000; 12(13): 965-968
5. G. Cheng, J. Chen, H. Ke, J. Shang and R. Chu. Synthesis, characterization and photocatalysis of $SnO_2$ nanorods with large aspect ratios. *Material Letters* 2011; 65(21-22): 3327-3329.
6. J. Huang, Z, Yin and O, Zheng. Applications of ZnO in organic and hybrid solar cells. *Energy and Environmental science* 2011; 4: 3861-3877.
7. Y. Zhang, M.K. Ram, E.K. Stefanakos and D.Y. Goswami. Synthesis, characterization and applications of ZnO nanowires. *Journal of nanomaterials* 2012; 2012: 1-22.
8. P. Samarasekara, A.G.K. Nisantha and A.S. Disanayake. High Photo-Voltage Zinc Oxide Thin Films Deposited by DC Sputtering. *Chinese Journal of Physics* 2002; 40(2): 196-199.
9. P. Samarasekara, N.U.S. Yapa, N.T.R.N. Kumara and M.V.K. Perera. $CO_2$ Gas Sensitivity of Sputtered Zinc Oxide Thin Films. *Bulletin of Materials Science* 2007; 30(2): 113-116.
10. P. Samarasekara. Liquid Junction Photocells Synthesized with Dye Coated Zinc Oxide Films. *Journal of Science of the University of Kelaniya Sri Lanka* 2010; 5: 25-31
11. P. Samarasekara and Udumbara Wijesinghe. Optical properties of spin coated Cu doped ZnO nanocomposite films. *GESJ:Physics* 2015; 2(14): 41-50.
12. P. Samarasekara, Udumbara Wijesinghe and E.N. Jayaweera. Impedance and electrical properties of Cu doped ZnO thin films. *GESJ:Physics* 2015; 1(13): 3-9.
13. P. Samarasekara, P. G. D. C. K. Karunarathna, B. M. C. M. Bandaranayake, J. S.T.

Wickramasinghe and C. A. N. Fernando. Spin coated multilayered CuO/ZnO films with high output power under irradiation. *Materials Research Express* 2018; 6(3): 036415.

14. P. Samarasekara and N.U.S. Yapa. Effect of sputtering conditions on the gas sensitivity of Copper Oxide thin films. *Sri Lankan Journal of Physics* 2007; 8: 21-27.

15. X. Zhang, F. Zhang and K.Y. Chan. Synthesis of titania-silica mixed oxide mesoporous materials, characterization and photocatalytic properties. *Applied Catalysis A: General* 2005; 284(1-2): 193-198.

16. D.G. Shchukin and R.A. Caruso. Template synthesis and photocatalytic properties of porous metal oxide spheres formed by nanoparticle infiltration. *Chemistry of Materials* 2004; 16: 2287-2292.

17. O. Dlugosz, K. Szostak and M. Banach. Photocatalytic properties of zirconium oxide-zinc oxide nanoparticles synthesized using microwave irradiation. *Applied Nanoscience* 2020; 10: 941-954.

18. C.W. Zou and W. Gao. Fabrication, optoelectronic and photocatalytic properties of some composite oxide nanostructures. *Transactions on electrical and electronic materials* 2010; 11(1): 1-10.

19. B. Pal, M. Sharon and G. Nogami. Preparation and characterization of $TiO_2/Fe_2O_3$ binary mixed oxides and its photocatalytic properties. *Materials Chemistry and Physics* 1999; 59(3): 254-261.